\documentclass[prd,twocolumn,nofootinbib]{revtex4}
\usepackage{graphicx, epsfig}
\usepackage{color}
\usepackage{mathrsfs}
\usepackage{bm}
\usepackage{amsmath,amssymb,yhmath}
\usepackage[caption=false]{subfig}
\usepackage{hyperref}
\usepackage[normalem]{ulem}
\usepackage{mathtools}

\newcommand{\be}{\begin{equation}}
\newcommand{\ee}{\end{equation}}
\newcommand{\ba}{\begin{eqnarray}}
\newcommand{\ea}{\end{eqnarray}}

\newcommand{\nn}{\nonumber}

\begin{document}
\title{Dynamical quantum collapse and an experimental test}
\author{Tanmay Vachaspati$^*$}
\affiliation{
$^*$Physics Department, Arizona State University, Tempe, Arizona 85287, USA. \\
}

\begin{abstract}
The quantum measurement problem may have a resolution in de Broglie-Bohm
theory in which measurements lead to dynamical wavefunction collapse. We study
the collapse in a simple setup and find that there may be slight differences between
probabilities derived from standard quantum mechanics versus those from 
de Broglie-Bohm theory in certain situations, possibly paving the way for an 
experimental test.
\end{abstract}

\maketitle




%

The quantum measurement problem has a long history, dating back to the discovery
of quantum mechanics, sometimes stated in terms of ``wavefunction collapse'' or
the ``projection postulate''. The essence of the problem is that the quantum state
is supposed to change instantaneously and in a stochastic, non-Hamiltonian way 
when a measurement is made. The collapse is even more puzzling because the
detector is built from microscopic constituents whose interactions with the quantum
system should be described by Hamiltonian evolution.

A variety of ideas have been proposed to resolve the quantum measurement problem
(see \cite{Bassi:2012bg} for a review). 
Here we will adapt the de Broglie-Bohm (dBB) idea as 
applied specifically to a non-relativistic, spinless, quantum
particle (for related references 
see~\cite{Bohm:1951xw,Bohm:1951xx,Holland:1993ee,Struyve:2004xd,Perez:2005gh,
Valentini:2006yj,Ryssens:2019kkm,Bassi:2003gd,Bassi:2004eu,Gasbarri:2017alx,
Nikolic:2012wj}).
In this approach, the 
particle is described by the wavefunction, $\psi(t,x)$, and a classical realization, $r(t)$. 
The wavefunction evolves according to the Schrodinger equation; the realization
evolves according to an equation of motion that we will specify below. The constituents
of the detector are quantum but there is a collective coordinate (e.g. center of mass
coordinate) that is well described by a classical variable and is the ``pointer'' variable.
The particle is ``detected'', by definition, once the wavefunction has collapsed on 
to the detector. We will describe the system in more detail in Sec.~\ref{system}.

Our strategy in this paper is to consider a quantum particle in a one-dimensional, 
infinite square
well in which we also place a detector that can detect the particle's position. 
Initially the particle is in its ground state and the rules of quantum mechanics (QM) 
tell us that the detection probability is given by the square of the wavefunction. 
Using the dBB theory we recover this result but by a dynamical process
in which we explicitly see the collapse of the wavefunction. Then we study the
situation when there are multiple detectors placed within the square well. In this
case, the rules of quantum mechanics would still lead to the usual detection probability
given by the square of the wavefunction. However wavefunction
collapse is a dynamical process in dBB theory and time-evolution in the presence of 
several detectors differs from that when there is only one detector. Roughly speaking,
it takes time for the wavefunction to collapse at the location of a detector and so
the probability density $|\psi(t,x)|^2$ changes with time and the probability for
a given detector to detect the particle is not simply given by the initial
probability density $|\psi (0,x)|^2$.
This result is sensitive to the classical nature of the detector -- to the number
of microscopic degrees of freedom that are represented in the collective coordinate -- 
and the differences between QM and dBB are most dramatic for small detectors. This 
leads to the possibility of experimentally distinguishing standard QM from dBB theory, 
and to a possible resolution of the quantum measurement problem.

\section{System}
\label{system}


\begin{figure}
      \includegraphics[width=0.4\textwidth,angle=0]{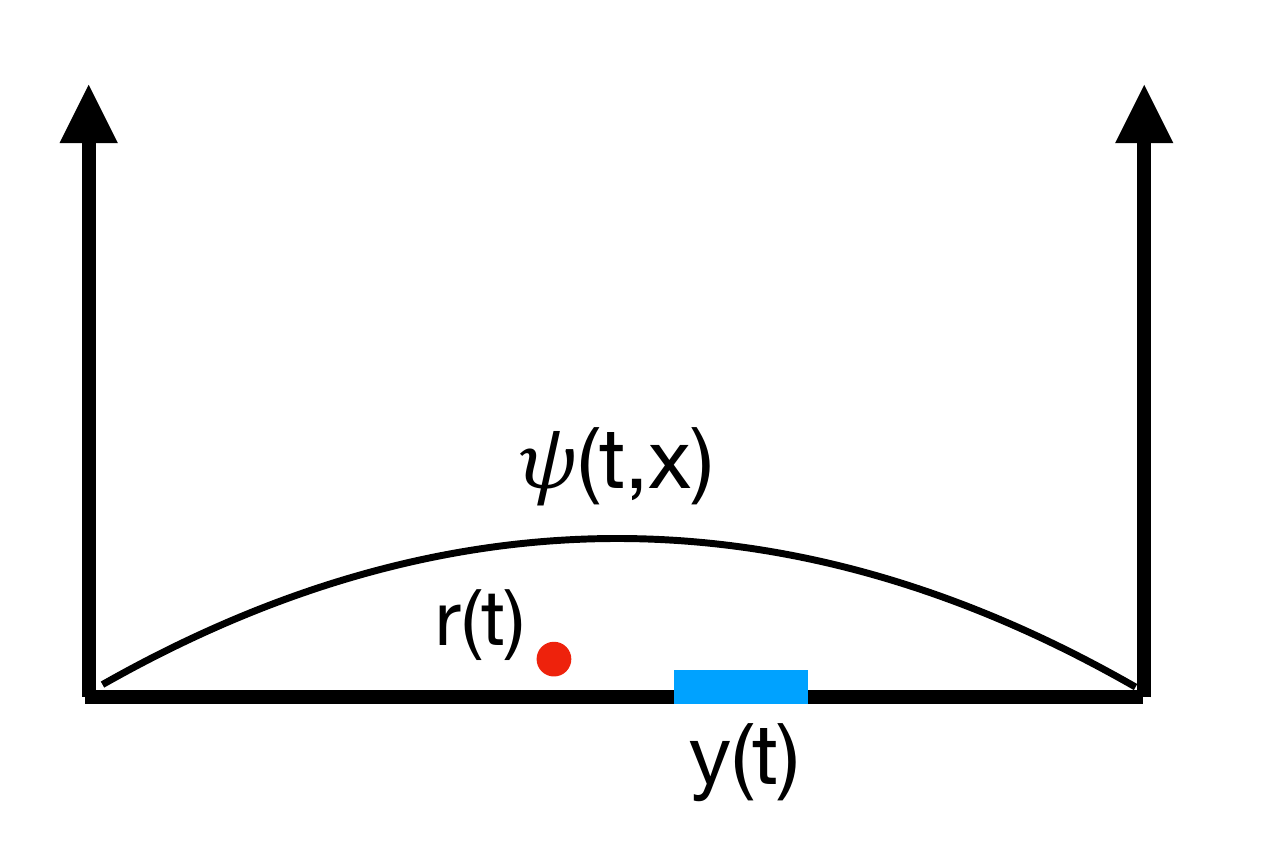}
  \caption{A sketch of the infinite square well with a particle and a single detector
  (represented by the blue horizontal bar). The variables in the dBB theory
  are the wavefunction $\psi(t,x)$ of the quantum particle, the particular realization
  $r(t)$ of the paricle, and the classical variable, $y(t)$, of the detector.}
  \label{squareWellDrawing}
\end{figure}

%
%
%
%

We consider a quantum particle with position variable $x$ and a
particle detector in a one-dimensional, infinite
square well as sketched in Figure~\ref{squareWellDrawing}. 
The detector
is made up of a number of microscopic degrees of freedom. Call
these $\xi_i$ where $i=1,\ldots,N$. Then the Hamiltonian can be written as
\ba
H &=& \frac{p_x^2}{2m} + V(x) 
+ \sum_i \frac{p_{\xi_i}^2}{2\mu} + U_1(\{|\xi_i-\xi_j|\})  \nn \\
&& + U_2\biggr (\sum_i \xi_i/N \biggr ) - \lambda W(x) \sum_i \xi_i
\ea
Here we use standard notation: $p_x$ and $p_{\xi_i}$ refer to conjugate
momenta for $x$ and $\xi_i$; $m$ and $\mu$ are masses of the quantum
particle and the detector microscopic degrees of freedom; $V(x)$ is the infinite square
well potential
\be
V(x) = \begin{cases} 0, & x\in [-L,L] \\ \infty, & {\rm otherwise} \end{cases}
\ee
$U_1(\{ |\xi_i-\xi_j|\})$ is some potential function for the relative internal coordinates of
the detector; $U_2$ is the potential for the collective coordinate of the detector,
taken to be
\be
y \equiv  \frac{1}{N} \sum_i \xi_i 
\ee 
The ``window function'' $W(x)$ 
defines 
the range of $x$ in which there is interaction with the detector. For example,
this might be a Gaussian function, $W(x) \propto \exp(-x^2/2\sigma^2)$. 
For simplicity, we will choose the window function to be a top-hat,
\be
W(x) = \begin{cases} 1 & x \in [x_0-d,x_0+d] \\ 0, & {\rm otherwise} \end{cases}
\label{tophat}
\ee
where $x_0$ is the central location of the detector and $2d$ is its full width.

The Hamiltonian can now be written as
\ba
H &=& \frac{p_x^2}{2m} + V(x)  \nn\\
&&+  \frac{p_y^2}{2N\mu} + U_2(y)
- \lambda N y W(x) + \ldots
\ea
where the ellipsis denote terms that describe the internal degrees of
freedom of the detector and will be ignored.

The detector collective coordinate $y$ is taken to be classical.
This is a valid approximation when the wavefunction for $y$ is in the
form of a tight wavepacket that remains tight throughout the evolution.
For a free particle the time-scale for the spreading of the wavepacket
is inversely proportional to the mass of the particle. So our classical
treatment of $y(t)$ should be valid for large $N\mu$. 
Then we may write,
\be
H = \frac{p_x^2}{2m} + V(x)  - \lambda N W(x) y(t) + H(y(t))
\ee

The Schrodinger equation can now be written as
\be
-\frac{1}{2m} \partial_x^2 \psi + V(x) \psi - \lambda N y(t) W(x) \psi
= i \partial_t \psi
\ee
while the equation for the classical variable $y$ is
\be
N\mu {\ddot y} + \partial_y U_2(y) = \lambda N W(x) 
\label{yeq}
\ee

The problem is that \eqref{yeq} involves the classical variable
$y$ and the quantum variable $x$ through $W(x)$. In the semiclassial
approximation $W(x)$ would be replaced by its expectation value.
Here however we will replace $x$ by a ``realization'' of the 
particle's position, $r(t)$. This modification -- that interactions occur 
between realizations -- is not discussed in dBB theory but seems like
a natural extension. So the $y$ equation becomes,
\be
N\mu {\ddot y} + \partial_y U_2 = \lambda N W(r(t)) 
\label{yeqr}
\ee
We now need an equation for the realization $r(t)$.

The probability distribution $|\psi(t=0,r)|^2$ is used to choose
the initial value $r(t=0)$. 
Subsequently $r(t)$ obeys the ``guidance equation'' of dBB theory
\cite{Bohm:1951xw,Bohm:1951xx},
\be
m {\dot r} = \partial_r S(t,r)
\label{mrdot}
\ee
where $S$ is the phase of the wavefunction: $\psi (t,x) = R(t,x) \exp(iS(t,x))$
where $R$ and $S$ are real. Note that we do not restrict $R$ to be non-negative.
The functions $R$ and $S$ will vary smoothly, even through points where $R=0$.



To summarize, the equations describing the quantum particle and the detector
are:
\be
-\frac{1}{2m} \partial_x^2 \psi + V(x) \psi - \lambda N y(t) W(x) \psi
= i \partial_t \psi
\ee
\be
\mu {\ddot y} + \frac{1}{N} \partial_y U_2(y) = \lambda W(r) 
\ee
\be
m {\dot r} = \partial_r S(t,r)
\ee
where $r(t=0)$ is drawn from the probability distribution $|\psi(t=0,r)|^2$.
The initial conditions for the detector variable are taken to be $y(0)=0={\dot y}(0)$.

We can generalize the model to several detectors by summing
over detector variables and window functions. For $D$ identical detectors, the
equations become,
\ba
-\frac{1}{2m} \partial_x^2 \psi + V \psi 
- \lambda N \sum_{i=1}^D y_i(t) W_i(x) \psi &=& i \partial_t \psi 
\label{psiequation}\\
\mu {\ddot y}_i  + \frac{1}{N} \partial_{y_i} U_2(\{y_i\}) = \lambda W_i(r) , &&
\label{yequation} \\
m {\dot r} = \partial_r S(t,r) &&
\label{requation}
\ea

The function $U_2$ determines the nature of the detector, for example,
how will the detector respond if there is no detection? Here we will consider
the case that the detector state does not change unless there is an interaction
with the particle realization and, for simplicity, set $U_2=0$. Then the
state of the detector changes only if $r(t)$ lies where the window 
function does not vanish. This behavior can easily be modified for example
by choosing the window function $W$ to be $+1$ if $r$ is within the detector
and $-1$ (not $0$) otherwise.
In addition we can assume different dynamics for the detector variable by
choosing different potential functions $U_2$, for example, a restoring 
force that brings $y$ back to zero when there is no interaction. We can
also include dissipation in the dynamics of the detector variables. The
exact detector dynamics will depend on the physical interpretation of
the detector variable $y(t)$ (current, pointer location, etc.) and its dynamics
in the specific detector of interest.




\section{Wavefunction collapse in a square well}
\label{squarewell}


We now solve the system of equations \eqref{psiequation}, \eqref{yequation}, 
and \eqref{requation} numerically by discretizing $x$ and using
Visscher's algorithm to solve Schrodinger's equation~\cite{doi:10.1063/1.168415}. 
We choose parameters 
as follows: the lattice spacing $dx=0.1$ and the time-step is $dt=dx^2/4$. Our 
grid size consists of 200 points and the physical half-width of the box is $L=10$. 
The detector half-width $d=1$, so $d=L/10$. We choose $\lambda=0.01$, $m=1$ 
and $\mu=1$ and will consider $N=1,\ldots,10$ with most of the runs with $N=1$.
(Larger values of $N$ gave numerical problems in certain cases.) 
We will locate the detector at 
lattice points $0,10,\ldots, 90$ equivalent to
$x=0,1,\ldots , 8, 9$. (Note that not all parameters are relevant. For
example, we can absorb the masses by rescaling the $x$ and $t$
coordinates.)
%
Initially the particle is taken to be in its ground state,
\be
\psi(0,x) = \frac{1}{\sqrt{L}} \sin \left ( \frac{\pi (x-L)}{2L} \right )
\label{psi0x}
\ee
and the detector is undisturbed,
\be
y(0)=0,\ \ {\dot y}(0)=0
\ee
The particle realization $r(0)$ is chosen in various places in the domain 
$(-L,+L)$ and will lead to different outcomes.

Since the quantum particle is in an {\it infinite} square well, we set: 
$\psi(t,|x| \ge L) =0$. One issue is that then the phase
$S$ is undefined for $|x| \ge L$ and boundary conditions must be imposed
on the particle realization once $r(t)$ becomes $\pm L$. We will assume
``absorptive'' boundary conditions: if $r(t)$ reaches within a lattice spacing of
the boundaries, we set the value of $r$ to the boundary value ($\pm L$).
The boundary conditions are only important for the detectors that lie 
adjacent to the walls of the infinite square well.

\subsection{$r(0)$ outside the detector}
\label{r0outside}

If the initial particle realization, $r(0)$, lies outside the detector, {\it i.e.} outside 
the interval $(x_0-d,x_0+d)$, Equation~\eqref{yequation} (with $U_2=0$) tells 
us that $y(t)=0$. Then the wavefunction retains the form in \eqref{psi0x} for 
all times, and $r(t)=r(0)$. This is the trivial solution in which the whole system 
is stationary.

As mentioned above, in our setup a ``no detection of a particle'' is
not equivalent to ``detection of no particle''. The former is equivalent to
no interaction between the particle and the detector; it is as if the
detector is not even turned on. The latter would require a change in the
detector corresponding to the lack of a particle.

\subsection{$r(0)$ inside the detector}
\label{r0inside}

\begin{figure}
      \includegraphics[width=0.45\textwidth,angle=0]{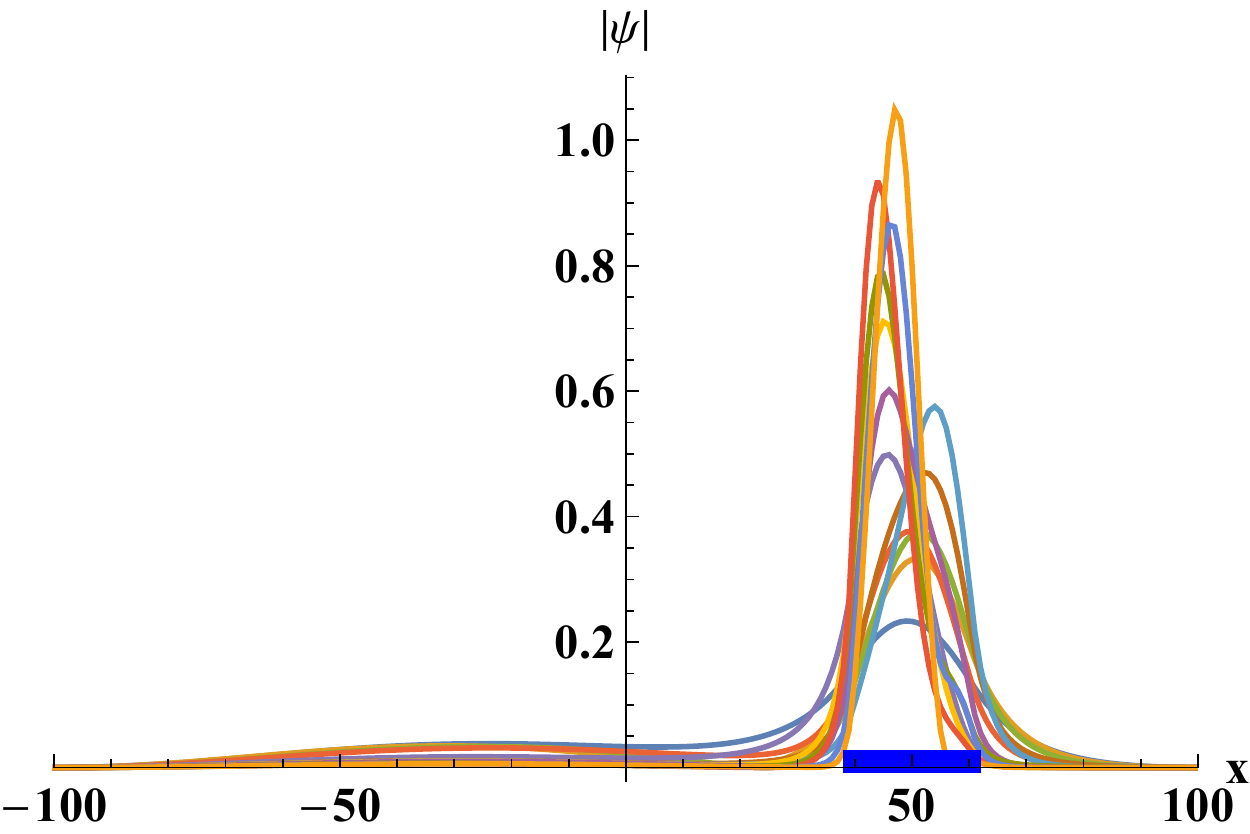}
  \caption{The collapse of the ground state wavefunction when the
  detector is located at $x=50$ and with half-width $d=10$ lattice units. 
  This run is for $N=1$ and with $r(0)=55$ lattice units. Other parameters 
  are as noted in the text. As time progresses, the wavefunction
  accumulates at the location of the detector. At time-step 289295, the
  probability for the particle to be in the detector reaches 0.95. With
  our chosen time-step, this corresponds to a time of collapse $t_c=723.238$ 
  in units of inverse particle mass.}
  \label{squareWellcollapse}
\end{figure}

If $r(0)$ lies within the detector, the evolution is non-trivial as seen in 
Figure~\ref{squareWellcollapse}. It is clear that the wavefunction collapses
onto the detector after some ``collapse time'' $t_c$. We define $t_c$ to
be the time taken for the probability of the particle to be in the detector
to become 0.95. 

\begin{figure}
      \includegraphics[width=0.45\textwidth,angle=0]{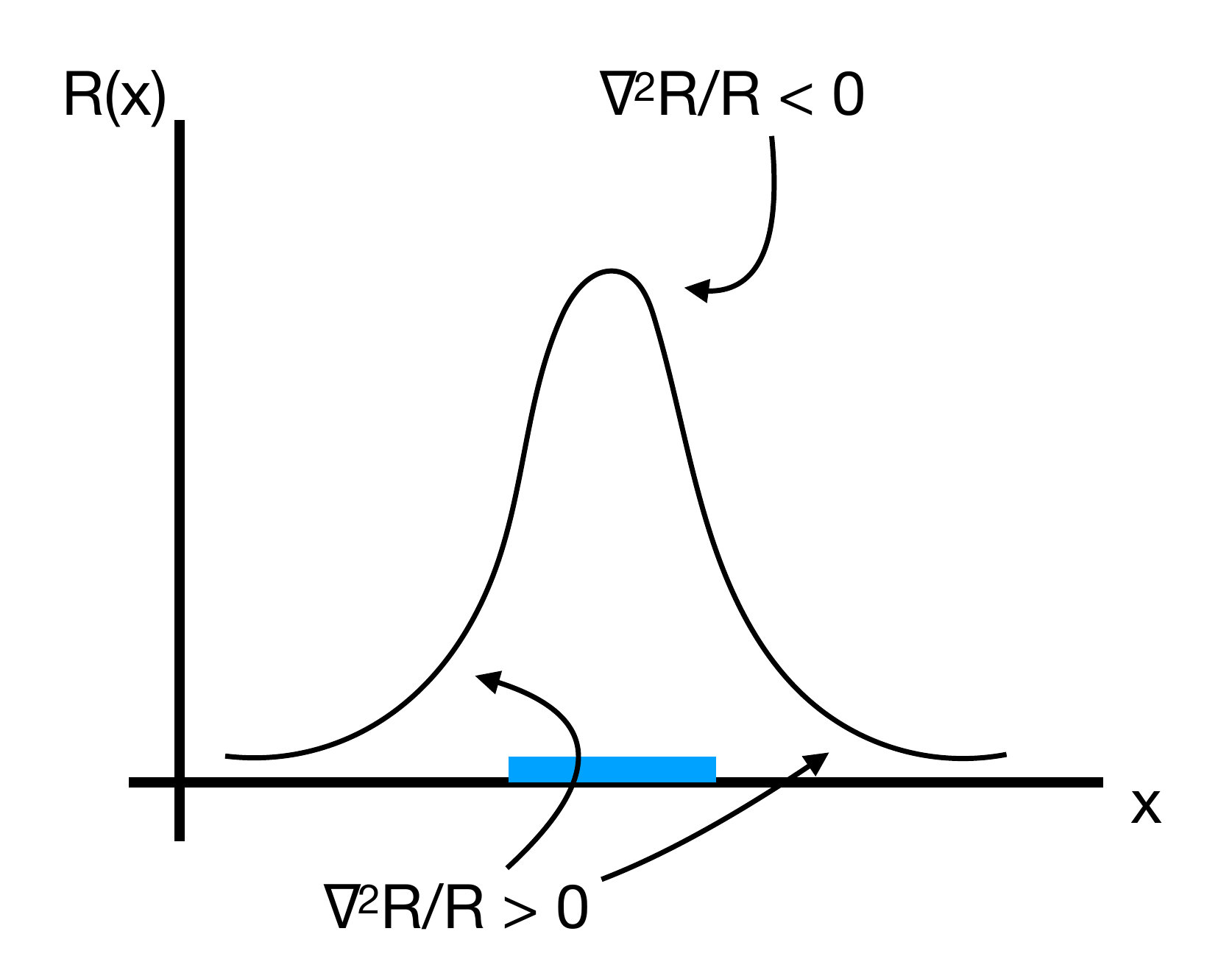}
  \caption{If the wavefunction has a peak within the detector -- the horizontal blue
  line in the sketch -- the quantum force 
  given by the last term in Eq.~\eqref{mddotr} tends
  to drive the particle realization out of the detector. This is seen by examing
  the signs of $\nabla^2R/R$ at various locations as indicated in the plot. Then
  the force, which is proportional to $\nabla(\nabla^2R/R)$ tends to push the realization
  out of the detector. This tendency must be overcome by the forces due to
  the detector potential for the wavefunction to collapse and the particle to
  be detected.}
  \label{detectorTrapplng}
\end{figure}

We can understand the tendency of the realization to stay within the
detector in the following way. If we write $\psi(t,x)=R(t,x) \exp(i S(t,x))$,
the equation for $S$ is,
\be
\partial_t S + \frac{(\nabla S)^2}{2m} + V_T (x) -
\frac{1}{2m} \frac{\nabla^2 R}{R} = 0
\ee
where $V_T$ represents all the potential terms (including for the detector).
The velocity of the realization is given by
\be
{\dot r} = \frac{\nabla S(x)}{m} \biggr |_{x=r(t)}
\ee
and the force on the realization is
\be
m {\ddot r} = - \nabla V_T (x) +
 \frac{1}{2m} \nabla \left ( \frac{\nabla^2R}{R} \right )
\label{mddotr}
\ee
The first term arises due to the potential well of the detector and tends to keep 
the realization within the detector. The second term acts in the opposite way
and tends to drive the $r(t)$ out of the detector. This is
illustrated in Figure~\ref{detectorTrapplng}. The term $-\nabla^2R/R$
acts like an additional potential term for the dynamics of $r(t)$. If there is a peak
of $R$ within the detector, $-\nabla^2R/R > 0$ at the peak and 
$-\nabla^2R/R < 0$ away from the peak. Hence the additional potential
term has a barrier at the peak of the wavefunction and a trough away
from the peak. Thus it provides a force on $r(t)$ towards the edges of the detector.
If this force can overcome the force due to the detector potential well $V_T$, the
realization may move out of the detector, leading to the possibility that
wavefunction collapse may only occur for ``strong detectors''.

\begin{figure}
      \includegraphics[width=0.45\textwidth,angle=0]{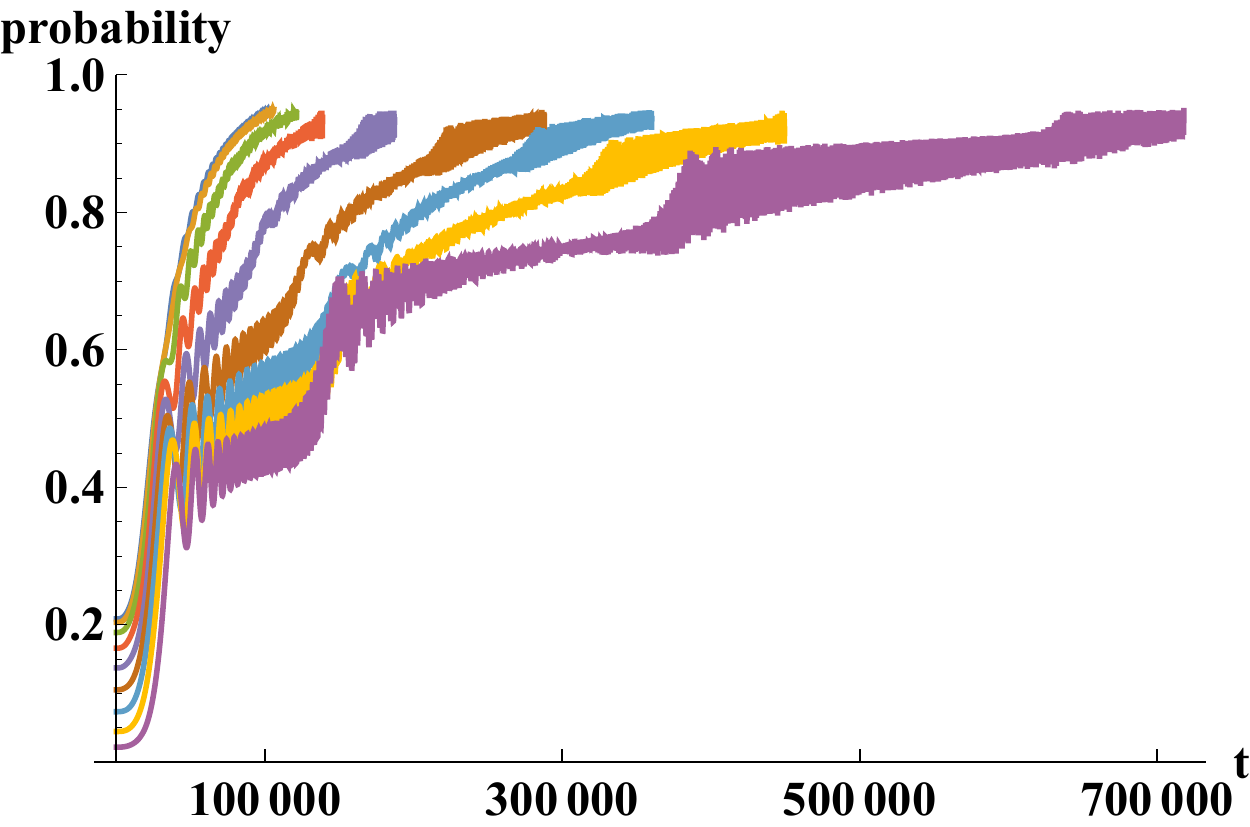}
  \caption{Plot of the probability to be in the detector versus time-step for various
  locations of the detector corresponding to Fig.~\ref{rzvstime}.
  We choose $\lambda=0.01$, $N=1$ and the
  detector locations range from $x_0=0,10,\ldots,80$, with $r(0)=5,15,\ldots,85$
  lattice units respectively. The collapse occurs most quickly for detectors
  placed close to the center of the square well and takes longer for detectors
  close to the boundaries. The detector at $x_0=90$ lattice units is not shown 
  because $r(t)$ quickly moves to the boundary of the square well; the curves
  for the $x_0=0$ and $x_0=10$ detectors lie too close to each other to be
  distinguishable in the plot.
  }
  \label{pvstime}
\end{figure}

The rate of collapse of the wavefunction depends on the location of the
detector. In Figure~\ref{pvstime} we plot the probability of the particle to
be in the detector as a function of time for various locations of the detector.
There are oscillations of the probability during the collapse that are most
prominent when the detector is placed close to the boundary of the infinite
square well.

\begin{figure}
      \includegraphics[width=0.45\textwidth,angle=0]{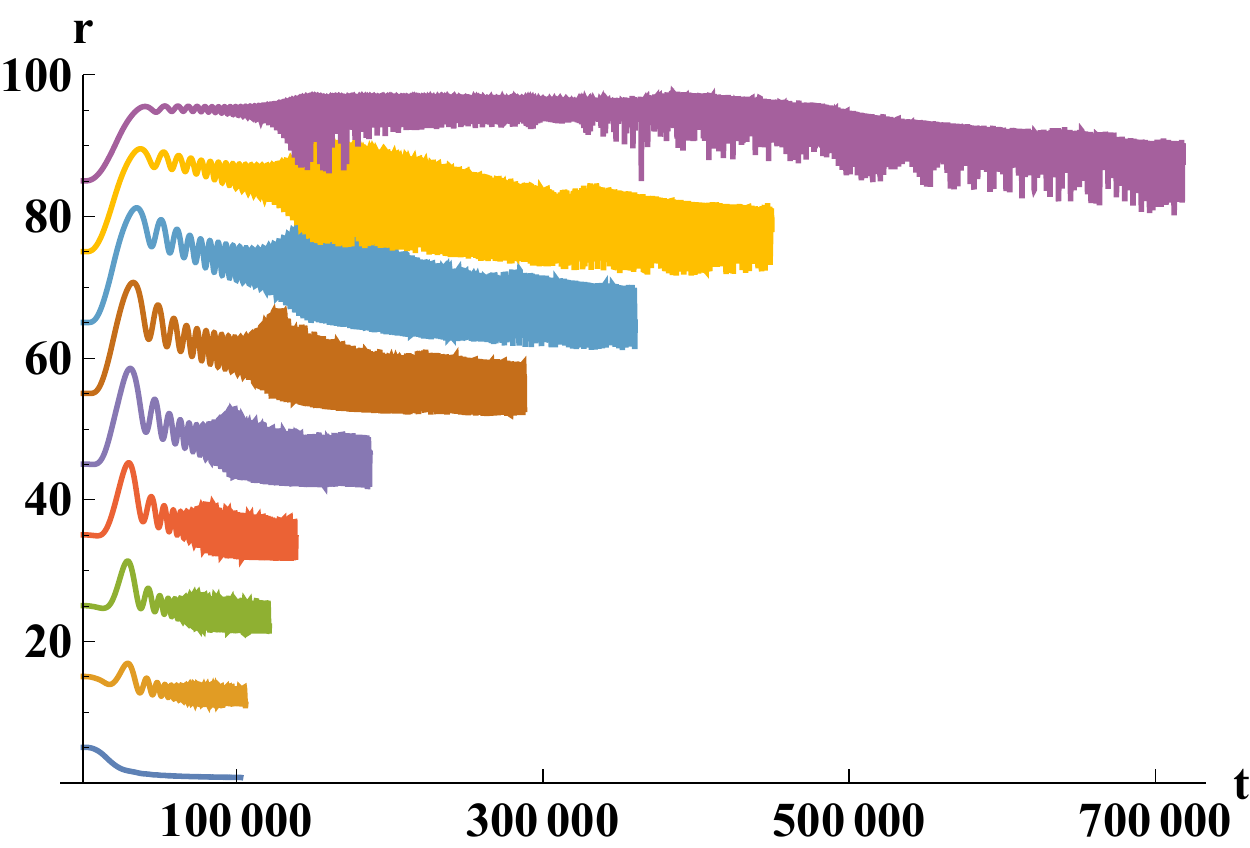}
  \caption{Plot of the realization $r(t)$ vs. time-step for the same parameters
  and initial conditions as in Fig.~\ref{pvstime}. For example, the fifth (purple) 
  curve from below 
  corresponds to a detector at $x_0=40$ with half-width $10$ lattice units and 
  with $r(0)=45$ lattice units. Although the particle realization is initially inside 
  the detector, it briefly strays out of the detector (to about $x=58$) but then 
  returns to stay trapped within the detector until full collapse occurs.
  }
  \label{rzvstime}
\end{figure}

We now turn to the evolution of the particle realization, $r(t)$. This is
plotted in Figure~\ref{rzvstime} for various locations of the detector. Here
too we see rapid oscillations in $r(t)$ but the most interesting feature
is that $r(t)$ strays out of the detector in which it started but then returns
to the detector. This will have important consequences when we
consider multiple detectors as in the sections below.

\begin{figure}
      \includegraphics[width=0.45\textwidth,angle=0]{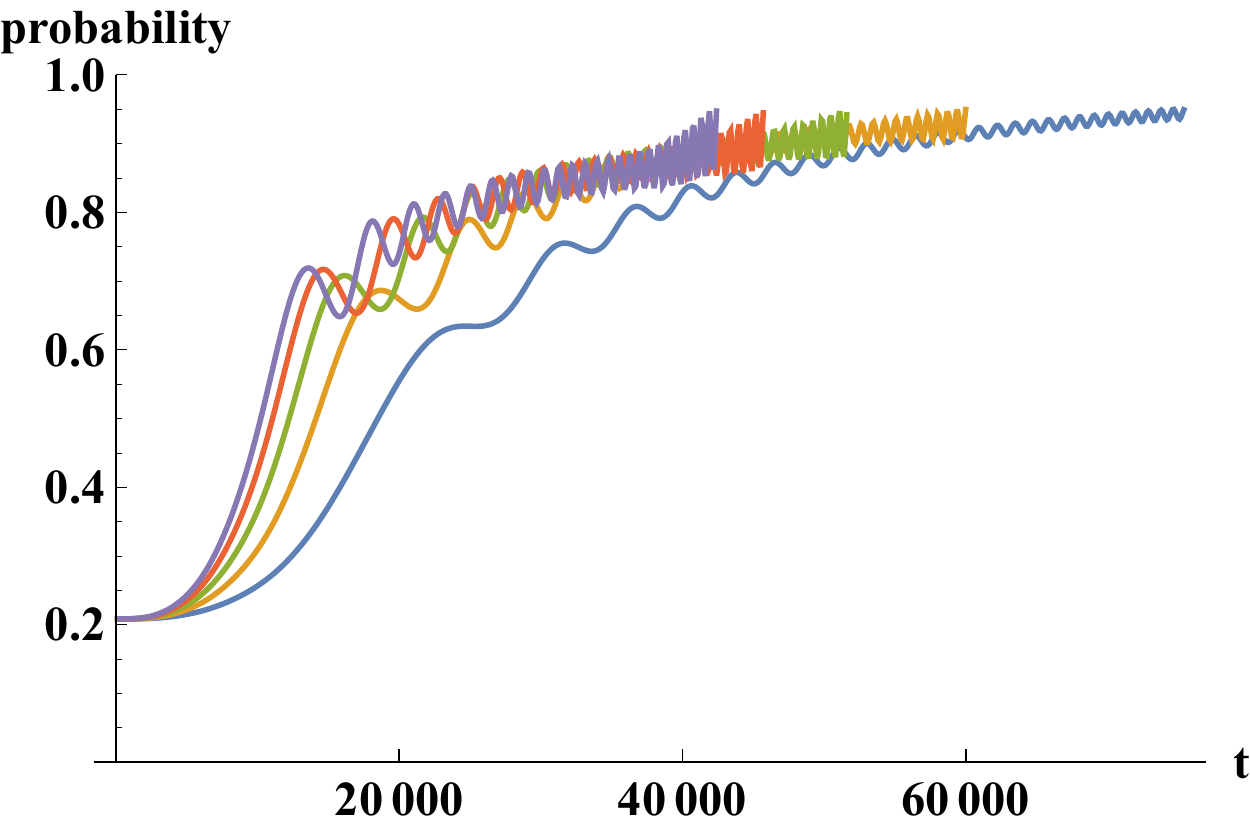}
  \caption{Plot of the probability to be in the detector versus time-step for 
  $N=2$ (blue), 4 (orange), 6 (green), 8 (red), and 10 (purple), for
  $\lambda=0.01$ when the detector is placed at the
  center of the square well and $r(0)=5$ lattice units. As $N$ increases
  the collapse occurs more quickly.
   }
  \label{pvstimelambda01x00}
\end{figure}

\begin{figure}
      \includegraphics[width=0.45\textwidth,angle=0]{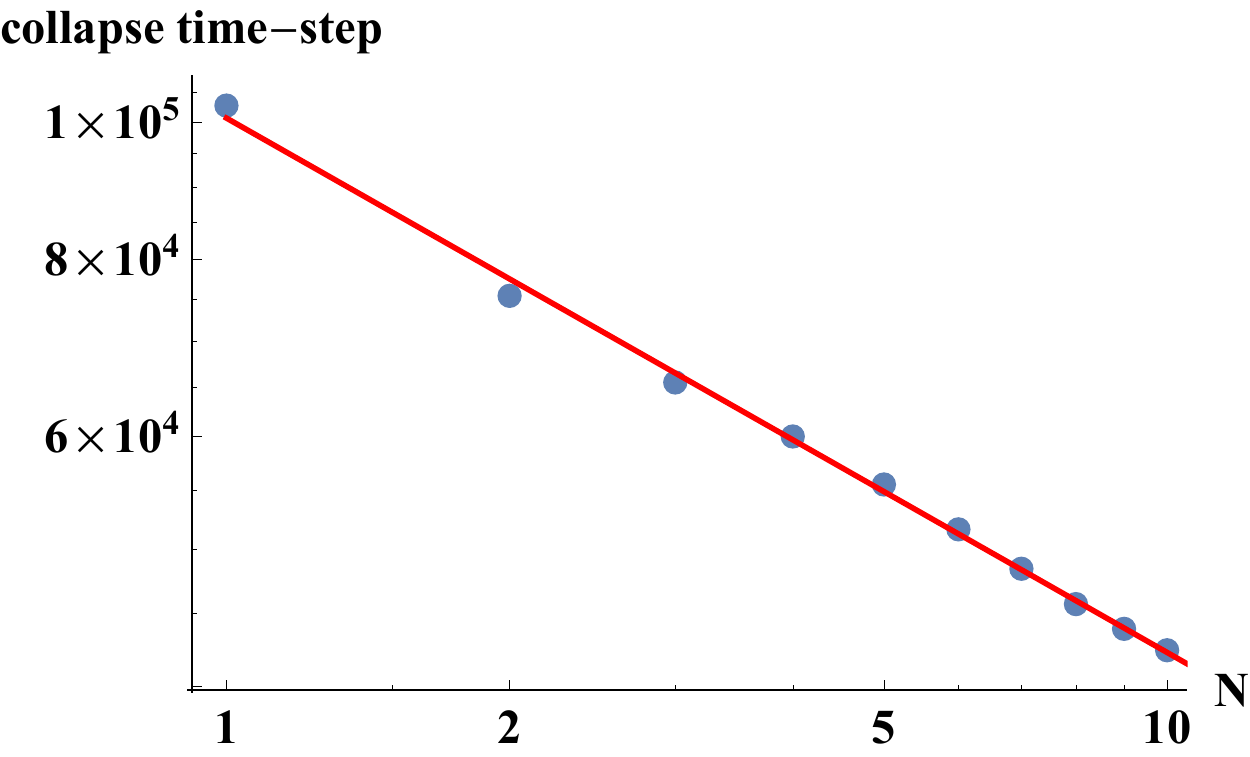}
  \caption{Log-log plot of the time of collapse when the probability to be in the detector 
  reaches 0.95 versus $N$ for
  $\lambda=0.01$ when the detector is placed at the
  center of the square well and $r(0)=5$ lattice units. The best fit line
  is also shown: $y =  11.52 - 0.38 x$ and corresponds to an $N^{-0.38}$
  fall off of the collapse time with increasing number of degrees of freedom.
   }
  \label{collapsetimeVsndof}
\end{figure}

Next we consider the dependence of the collapse on the number of
degrees of freedom of the detector, $N$. We fix the position of the
detector to be $x_0=0$ lattice units and study wavefunction collapse for
$N=2,4,6,8,10$ as shown in Figure~\ref{pvstimelambda01x00}. Increasing
$N$ makes the collapse occur more quickly. We quantify this feature
in Figure~\ref{collapsetimeVsndof} where we plot the collapse time $t_c$
versus $N$ on a log-log scale. The straight line indicates a power law,
\be
t_c \propto N^{-0.38}
\ee
This shows that detectors with larger number of internal degrees of freedom 
make the wavefunction collapse more quickly, in accordance with expectations.

%

%
%
%

\subsection{Stochastic $r(0)$}
\label{stochastic}

The initial realization, $r(0)$, is a stochastic variable. From the analysis above,
if $r(0)$ is inside the detector, then the wavefunction will collapse on to the
detector; while if $r(0)$ is outside the detector, then the wavefunction does not
collapse and the particle is not detected. 
Therefore the dBB theory with a 
single detector will agree with quantum mechanics predictions if
the probability density for $r(0)$, denoted $P$, is given by the square
of the wavefunction,
\be
P[r(0)=r_0] = |\psi(0,r_0)|^2
\label{r0prob}
\ee
This conclusion applies when the initial wavefunction is in a
stationary state. It is not clear to us what distribution of $r(0)$ is
appropriate when the initial state is not a stationary state.

\section{Two detectors}
\label{twodetectors}

\begin{figure}
      \includegraphics[width=0.22\textwidth,angle=0]{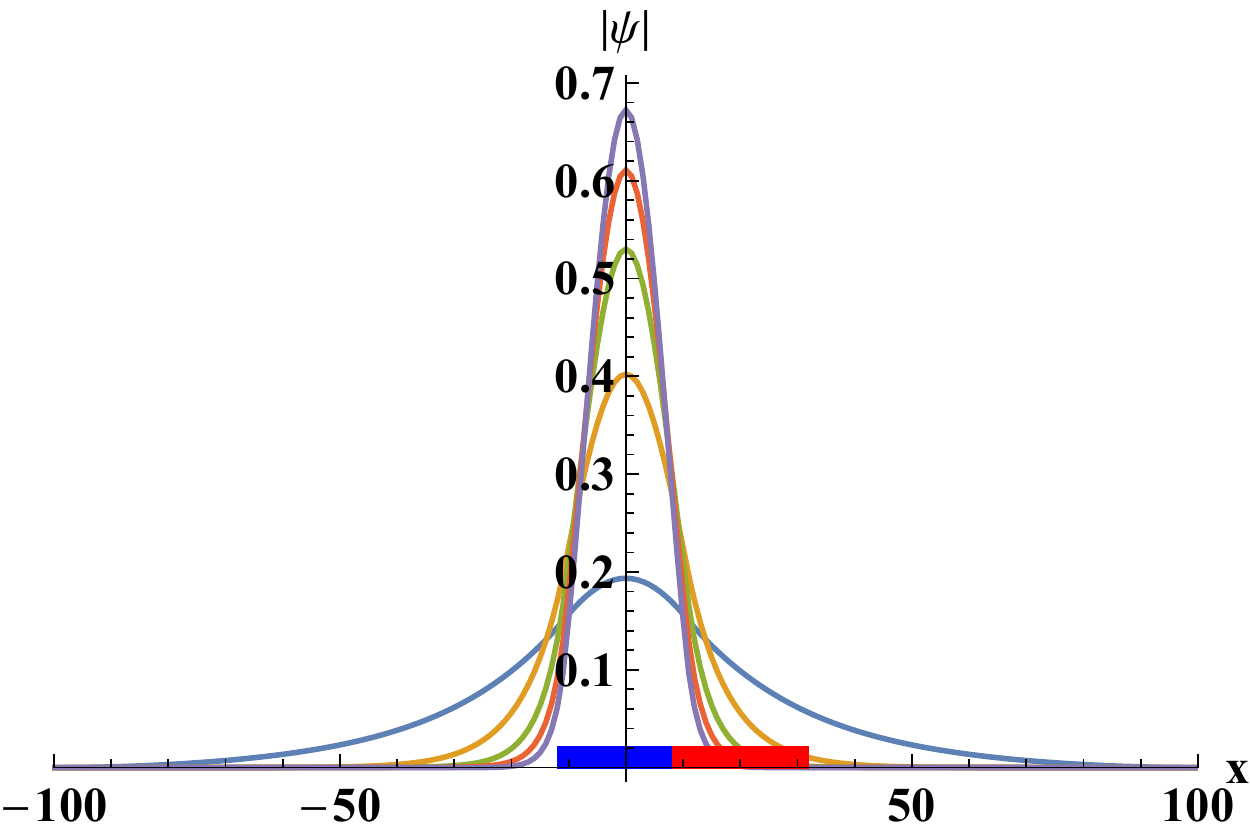}
      \includegraphics[width=0.22\textwidth,angle=0]{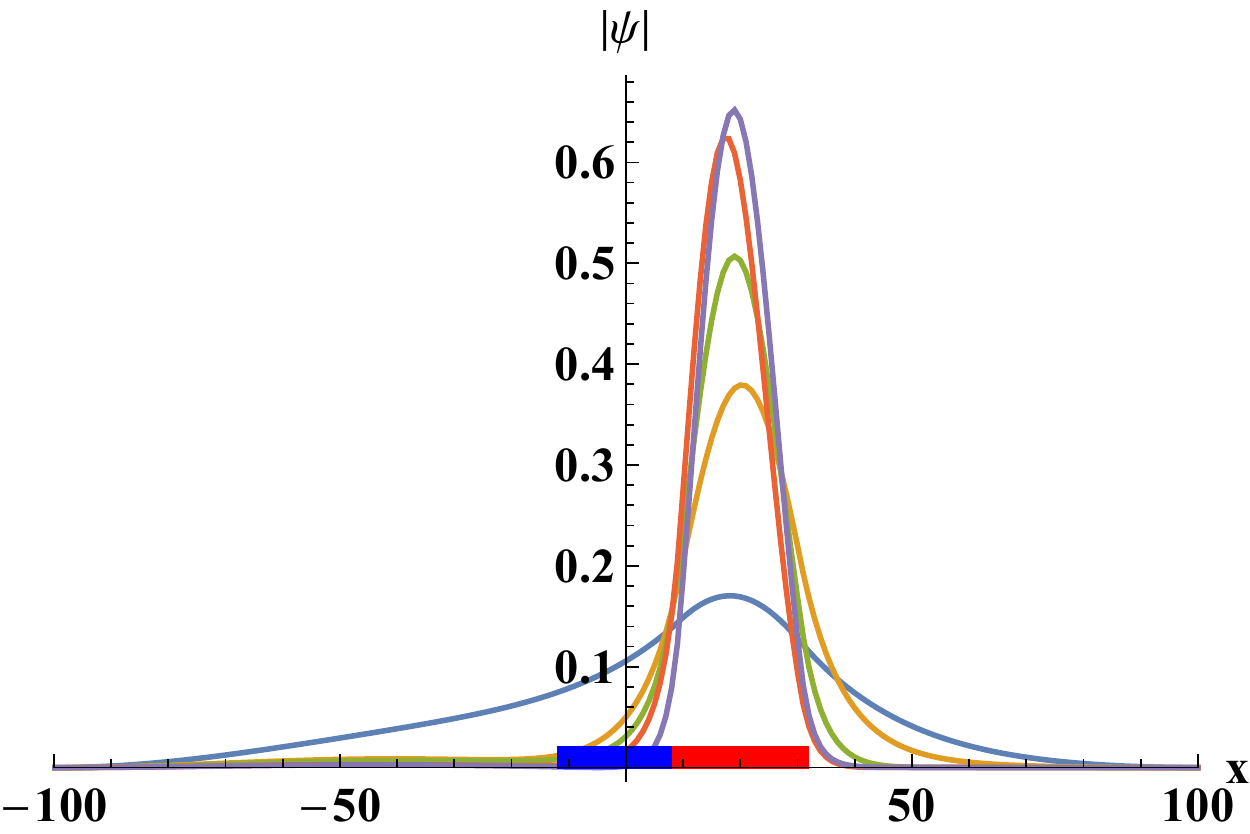}
      \includegraphics[width=0.22\textwidth,angle=0]{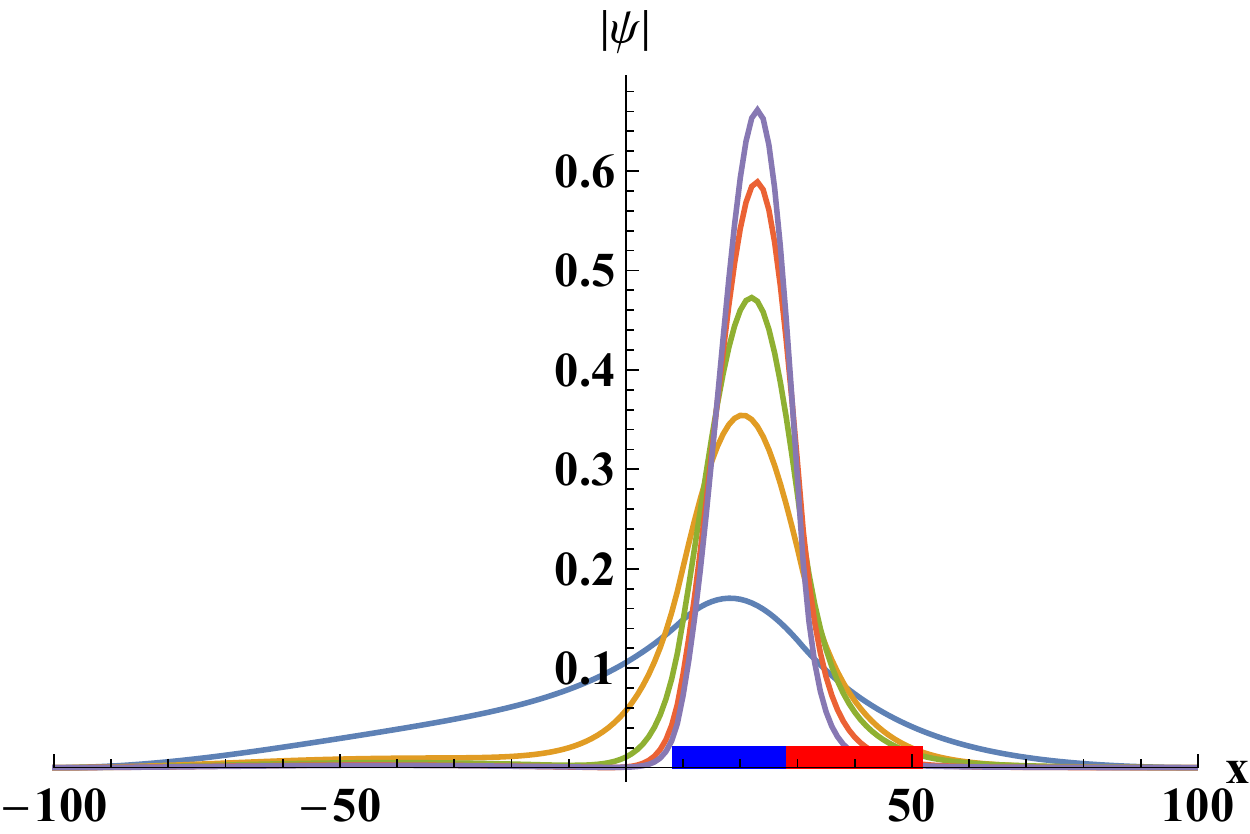}
      \includegraphics[width=0.22\textwidth,angle=0]{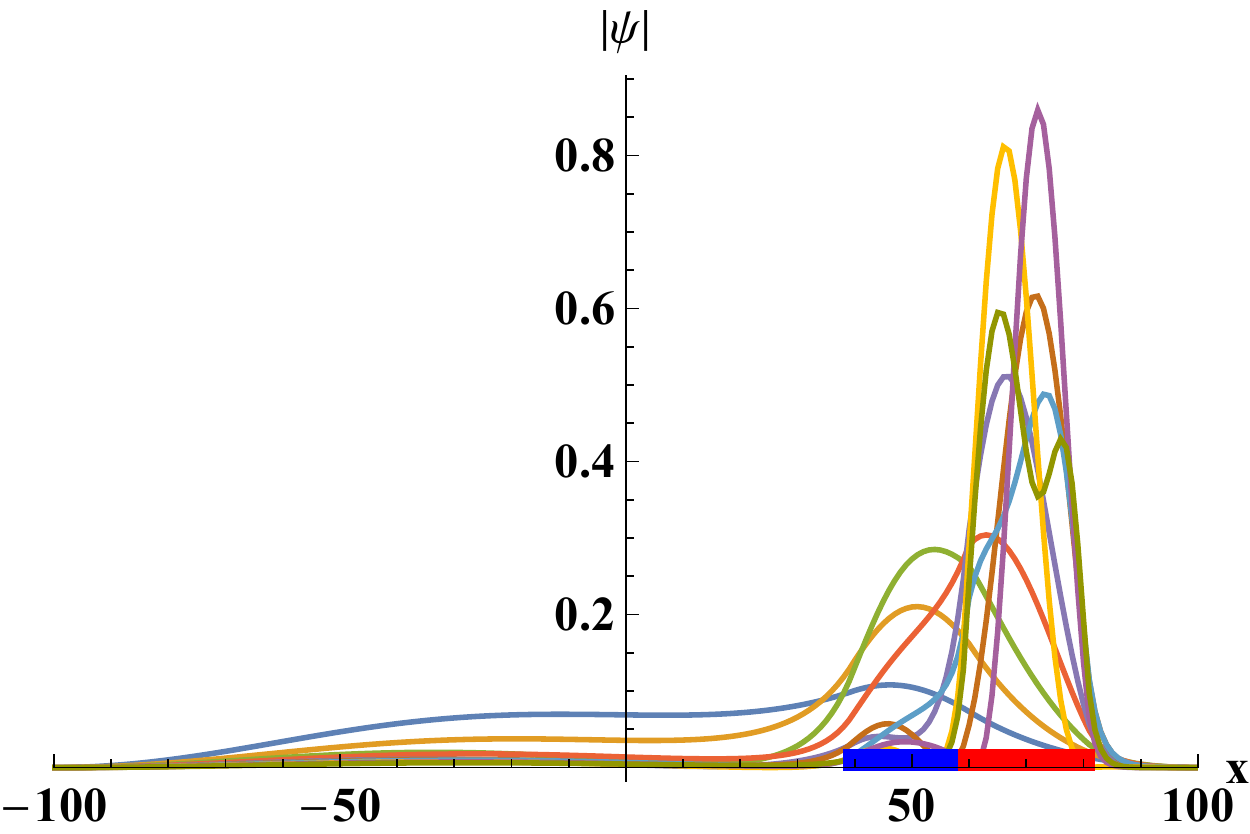}
  \caption{
    Wavefunction collapse with two detectors in various locations shown by thick 
    blue and red horizontal line segments. (If the detectors are widely separated,
    the system evolves as in the single detector case.) The model parameters
    are $\lambda=0.01$ and $N=1$. In the first three cases, the second detector 
    plays no role and the collapse proceeds as if there was only one detector.
    In the fourth case (bottom right) $r(0)$ is chosen within the left detector,
    which is also where the wavefunction starts to collapse. However, after some
    time the collapse shifts to the detector on the right where the particle is
    eventually detected.
   }
  \label{psiVstime1det02det10rz05}
\end{figure}

We now consider wavefunction collapse when there are two detectors
simultaneously present. The collapse should occur on only one detector 
and this is indeed what happens as can be seen in 
Fig.~\ref{psiVstime1det02det10rz05}. With more than one detector, the
evolution of the wavefunction can be affected by the presence of the
``dormant'' detectors, leading to new dynamics and differences with
standard quantum mechanics. For example, if the particle is realized in one
detector but meanders into another detector, as seen in 
Fig.~\ref{rzvstime}, the collapse may occur on
the second detector as shown in the bottom right plot of 
Fig.~\ref{psiVstime1det02det10rz05}.

%
%

\section{Detector array}
\label{detarray}

Now we consider the situation where the square well is lined with an array of 
10 detectors that cover $x\in [-L,L]$. 
We solve equations \eqref{psiequation}, \eqref{yequation}, and \eqref{requation}
as before, scanning over $r(0)$, the initial value of the realization. In each run, the 
wavefunction collapses on a detector but the collapse onto a detector is affected by 
the presence of other detectors as we have already seen in the case of two
detectors in Sec.~\ref{twodetectors}. 

\begin{figure}
      \includegraphics[width=0.45\textwidth,angle=0]{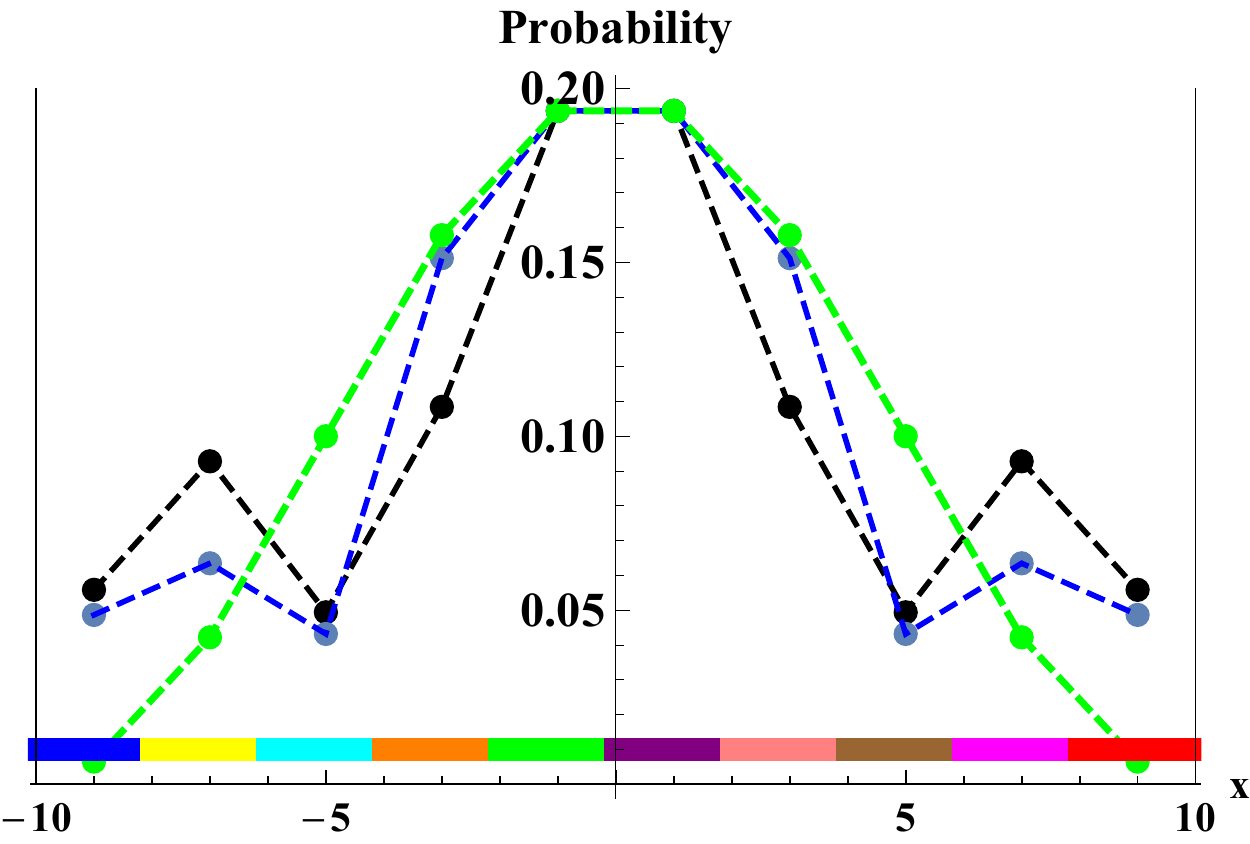}
  \caption{The probability for wavefunction collapse on a detector when there
  is an array of 10 detectors laid out across the infinite square well. The dBB theory
  result is shown in black dots joined by the dashed black curve for $N=1$ and
  by the blue dots and blue dashed curve for $N=2$, though with a lower
  collapse probability threshold of 0.75 (instead of 0.95) for $N=2$ case for numerical 
  reasons. The
  probabilities calculated in standard quantum mechanics are shown by the green 
  dots joined by the dashed green curve. The ten detectors are represented by the 
  thick horizontal colored lines.}
  \label{pplot}
\end{figure}

Next we factor in the probability distribution of $r(0)$ as in Eq.~\eqref{r0prob}.
Thus each initial value of $r(0)$ carries a weight given by the initial probability 
$|\psi(t=0,r(0))|^2$ and we can determine the probability that the wavefunction 
will collapse on any given detector. Define $S_n$ to be the set of $r(0)$ values
for which the wavefunction collapses on to the $n^{th}$ detector. Then the
probability for collapse on to the $n^{\rm th}$ detector is,
\be
p_n = \sum_{r_0 \in S_n} P[r_0]
\ee
The result for $p_n$ for each detector is shown in Figure~\ref{pplot} where 
we compare it to the usual quantum mechanics prediction,
\be
p_n^{(0)} = \int_{D_n} dx\, |\psi(0,x)|^2
\ee
where the integration is over the extent of the $n^{th}$ detector. We see 
from Figure~\ref{pplot} that
$p_n$ is different from $p_n^{(0)}$ 
for $N=1$ and $N=2$. (For the $N=2$
case, the time-dependent potential in \eqref{psiequation} is larger and we
encountered numerical issues for a handful of values of $r(0)$. So we
made our criterion defining collapse less stringent by 
setting the cutoff probability to count as a collapse to be 0.75 instead of 0.95.)
From Figure~\ref{pplot} we see that only the probabilities for the central two detectors
in dBB theory agree with quantum mechanics for $N=1$. When we increase
$N$ to 2, the agreement improves to the central four detectors. From 
this trend it appears that probabilities in dBB theory tend towards standard 
quantum mechanics probabilities as $N$ gets larger and will agree perfectly
in the $N\to \infty$ limit.
The different probabilities predicted by dBB theory and standard quantum
mechanics for smaller values of $N$ may pave the way for an experimental 
test of dBB theory.


\section{Conclusions}
\label{conclusions}

In quantum mechanics definite classical realizations of quantum systems are postulated 
when measurements are made but the quantum to classical transition (wavefunction
collapse) is not resolved. In dBB theory, there is
a dynamical resolution of wavefunction collapse. We have examined this process in a 
very specific simple setup. We find dynamical collapse of the wavefunction during a
measurement and that the collapse occurs more rapidly with larger number of degrees 
of freedom of the detector: the collapse time decreases with $N$ as $N^{-0.38}$ as in
Figure~\ref{collapsetimeVsndof}.
If the system consists of a single detector, we can postulate a probability distribution of 
the hidden variable of dBB theory such that the detection probabilities agree with those 
of standard quantum mechanics. With multiple detectors present, the
detection probabilities in dBB theory show departures from standard quantum
mechanics as in Figure~\ref{pplot}. 

The departures from standard quantum mechanics should be a general conclusion for all 
models in which there
is dynamical wavefunction collapse because the wavefunction has to ``decide'' on
which detector to collapse and this can take time during which the dynamics of
the wavefunction can be altered. (Departures from quantum mechanics predictions
in cosmology have also been studied in a Continuous Collapse
Localization model~\cite{Martin:2019jye,Martin:2019oqq}.)
With dynamical collapse and multiple measuring devices that are each trying to 
collapse the wavefunction, the detectors can start interfering with each other.
We expect that the departures from standard quantum mechanics will become minimal 
as we go to larger and more complex detectors. An experimental test of dBB
theory would be to obtain detection probabilities when there are multiple, small
detectors present in the experimental setup.

Our analysis has been restricted to a very simple and idealized system. It is
important to extend the analysis to more realistic systems. It would also be
useful to apply a similar analysis to discrete (spin) systems and to see if they
can provide an experimental test of dBB theory. Finally, it remains to be
seen if an analogous system can be set up in field theory in terms of a 
wavefunctional and a classical realization of the field.

\acknowledgements
I am grateful to George Zahariade for many illuminating discussions, to 
Ayush Saurabh for numerical help, and to Vachaspati for introducing me to David 
Bohm's ``Quantum Mechanics''.  I thank Paul Davies, Phil Tee, Vincent Vennin 
and Alex Vilenkin for comments.
TV is supported by the U.S.  Department of Energy, Office of High Energy Physics, 
under Award No.  DE-SC0018330 at Arizona State University.

\bibstyle{aps}
\bibliography{paper}

\begin{thebibliography}{15}
\expandafter\ifx\csname natexlab\endcsname\relax\def\natexlab#1{#1}\fi
\expandafter\ifx\csname bibnamefont\endcsname\relax
  \def\bibnamefont#1{#1}\fi
\expandafter\ifx\csname bibfnamefont\endcsname\relax
  \def\bibfnamefont#1{#1}\fi
\expandafter\ifx\csname citenamefont\endcsname\relax
  \def\citenamefont#1{#1}\fi
\expandafter\ifx\csname url\endcsname\relax
  \def\url#1{\texttt{#1}}\fi
\expandafter\ifx\csname urlprefix\endcsname\relax\def\urlprefix{URL }\fi
\providecommand{\bibinfo}[2]{#2}
\providecommand{\eprint}[2][]{\url{#2}}

\bibitem[{\citenamefont{Bassi et~al.}(2013)\citenamefont{Bassi, Lochan, Satin,
  Singh, and Ulbricht}}]{Bassi:2012bg}
\bibinfo{author}{\bibfnamefont{A.}~\bibnamefont{Bassi}},
  \bibinfo{author}{\bibfnamefont{K.}~\bibnamefont{Lochan}},
  \bibinfo{author}{\bibfnamefont{S.}~\bibnamefont{Satin}},
  \bibinfo{author}{\bibfnamefont{T.~P.} \bibnamefont{Singh}}, \bibnamefont{and}
  \bibinfo{author}{\bibfnamefont{H.}~\bibnamefont{Ulbricht}},
  \bibinfo{journal}{Rev. Mod. Phys.} \textbf{\bibinfo{volume}{85}},
  \bibinfo{pages}{471} (\bibinfo{year}{2013}), \eprint{1204.4325}.

\bibitem[{\citenamefont{Bohm}(1952{\natexlab{a}})}]{Bohm:1951xw}
\bibinfo{author}{\bibfnamefont{D.}~\bibnamefont{Bohm}}, \bibinfo{journal}{Phys.
  Rev.} \textbf{\bibinfo{volume}{85}}, \bibinfo{pages}{166}
  (\bibinfo{year}{1952}{\natexlab{a}}).

\bibitem[{\citenamefont{Bohm}(1952{\natexlab{b}})}]{Bohm:1951xx}
\bibinfo{author}{\bibfnamefont{D.}~\bibnamefont{Bohm}}, \bibinfo{journal}{Phys.
  Rev.} \textbf{\bibinfo{volume}{85}}, \bibinfo{pages}{180}
  (\bibinfo{year}{1952}{\natexlab{b}}).

\bibitem[{\citenamefont{Holland}(1993)}]{Holland:1993ee}
\bibinfo{author}{\bibfnamefont{P.~R.} \bibnamefont{Holland}},
  \bibinfo{journal}{Phys. Rept.} \textbf{\bibinfo{volume}{224}},
  \bibinfo{pages}{95} (\bibinfo{year}{1993}).

\bibitem[{\citenamefont{Struyve}(2004)}]{Struyve:2004xd}
\bibinfo{author}{\bibfnamefont{W.}~\bibnamefont{Struyve}}, Ph.D. thesis,
  \bibinfo{school}{Gent U.} (\bibinfo{year}{2004}), \eprint{quant-ph/0506243}.

\bibitem[{\citenamefont{Perez et~al.}(2006)\citenamefont{Perez, Sahlmann, and
  Sudarsky}}]{Perez:2005gh}
\bibinfo{author}{\bibfnamefont{A.}~\bibnamefont{Perez}},
  \bibinfo{author}{\bibfnamefont{H.}~\bibnamefont{Sahlmann}}, \bibnamefont{and}
  \bibinfo{author}{\bibfnamefont{D.}~\bibnamefont{Sudarsky}},
  \bibinfo{journal}{Class. Quant. Grav.} \textbf{\bibinfo{volume}{23}},
  \bibinfo{pages}{2317} (\bibinfo{year}{2006}), \eprint{gr-qc/0508100}.

\bibitem[{\citenamefont{Valentini}(2007)}]{Valentini:2006yj}
\bibinfo{author}{\bibfnamefont{A.}~\bibnamefont{Valentini}},
  \bibinfo{journal}{J. Phys.} \textbf{\bibinfo{volume}{A40}},
  \bibinfo{pages}{3285} (\bibinfo{year}{2007}), \eprint{hep-th/0610032}.

\bibitem[{\citenamefont{Ryssens}(2019)}]{Ryssens:2019kkm}
\bibinfo{author}{\bibfnamefont{W.}~\bibnamefont{Ryssens}}, Ph.D. thesis,
  \bibinfo{school}{KU Leuven, Dept. Phys. Astron.} (\bibinfo{year}{2019}),
  \eprint{1907.00258}.

\bibitem[{\citenamefont{Bassi and Ghirardi}(2003)}]{Bassi:2003gd}
\bibinfo{author}{\bibfnamefont{A.}~\bibnamefont{Bassi}} \bibnamefont{and}
  \bibinfo{author}{\bibfnamefont{G.~C.} \bibnamefont{Ghirardi}},
  \bibinfo{journal}{Phys. Rept.} \textbf{\bibinfo{volume}{379}},
  \bibinfo{pages}{257} (\bibinfo{year}{2003}), \eprint{quant-ph/0302164}.

\bibitem[{\citenamefont{Bassi et~al.}(2005)\citenamefont{Bassi, Ippoliti, and
  Adler}}]{Bassi:2004eu}
\bibinfo{author}{\bibfnamefont{A.}~\bibnamefont{Bassi}},
  \bibinfo{author}{\bibfnamefont{E.}~\bibnamefont{Ippoliti}}, \bibnamefont{and}
  \bibinfo{author}{\bibfnamefont{S.~L.} \bibnamefont{Adler}},
  \bibinfo{journal}{Phys. Rev. Lett.} \textbf{\bibinfo{volume}{94}},
  \bibinfo{pages}{030401} (\bibinfo{year}{2005}), \eprint{quant-ph/0406108}.

\bibitem[{\citenamefont{Gasbarri et~al.}(2017)\citenamefont{Gasbarri, Donadi,
  Bassi, and Toroš}}]{Gasbarri:2017alx}
\bibinfo{author}{\bibfnamefont{G.}~\bibnamefont{Gasbarri}},
  \bibinfo{author}{\bibfnamefont{S.}~\bibnamefont{Donadi}},
  \bibinfo{author}{\bibfnamefont{A.}~\bibnamefont{Bassi}}, \bibnamefont{and}
  \bibinfo{author}{\bibfnamefont{M.}~\bibnamefont{Toroš}},
  \bibinfo{journal}{Phys. Rev.} \textbf{\bibinfo{volume}{D96}},
  \bibinfo{pages}{104013} (\bibinfo{year}{2017}), \eprint{1701.02236}.

\bibitem[{\citenamefont{Nikolic}(2012)}]{Nikolic:2012wj}
\bibinfo{author}{\bibfnamefont{H.}~\bibnamefont{Nikolic}}, in
  \emph{\bibinfo{booktitle}{Applied Bohmian Mechanics}} (\bibinfo{year}{2012}),
  pp. \bibinfo{pages}{455--506}, \eprint{1205.1992}.

\bibitem[{\citenamefont{Visscher}(1991)}]{doi:10.1063/1.168415}
\bibinfo{author}{\bibfnamefont{P.~B.} \bibnamefont{Visscher}},
  \bibinfo{journal}{Computers in Physics} \textbf{\bibinfo{volume}{5}},
  \bibinfo{pages}{596} (\bibinfo{year}{1991}),
  \eprint{https://aip.scitation.org/doi/pdf/10.1063/1.168415},
  \urlprefix\url{https://aip.scitation.org/doi/abs/10.1063/1.168415}.

\bibitem[{\citenamefont{Martin and
  Vennin}(2019{\natexlab{a}})}]{Martin:2019jye}
\bibinfo{author}{\bibfnamefont{J.}~\bibnamefont{Martin}} \bibnamefont{and}
  \bibinfo{author}{\bibfnamefont{V.}~\bibnamefont{Vennin}}
  (\bibinfo{year}{2019}{\natexlab{a}}), \eprint{1906.04405}.

\bibitem[{\citenamefont{Martin and
  Vennin}(2019{\natexlab{b}})}]{Martin:2019oqq}
\bibinfo{author}{\bibfnamefont{J.}~\bibnamefont{Martin}} \bibnamefont{and}
  \bibinfo{author}{\bibfnamefont{V.}~\bibnamefont{Vennin}}
  (\bibinfo{year}{2019}{\natexlab{b}}), \eprint{1912.07429}.

\end{thebibliography}

\end{document}